\documentclass[12pt]{arXiv_class}

\title{\textbf{$\ell_1$ Adaptive Trend Filter via Fast Coordinate Descent}} 
\author{Mario Souto, Joaquim D. Garcia and Gustavo C. Amaral}

\begin{document}
\maketitle

\begin{abstract}
Identifying the unknown underlying trend of a given noisy signal is extremely useful for a wide range of applications. The number of potential trends might be exponential, thus their search  can be computationally intractable even for short signals. Another challenge is the presence of abrupt changes and outliers at unknown times which impart resourceful information regarding the signal's characteristics. In this paper, we present the $\ell_1$ Adaptive Trend Filter, which can consistently identify the components in the underlying trend and multiple level-shifts, even in the presence of outliers. Additionally, an enhanced coordinate descent algorithm which exploits the filter design is presented. Some implementation details are discussed and a version in the Julia language is presented along with two distinct applications to illustrate the filter's potential.
\end{abstract}

\section{Introduction}

Estimating the underlying dynamics of a signal of interest is an important task due to numerous reasons, for instance: denoising, forecasting, filtering and even more sophisticated analysis. No matter the goal, determining the constituting components of the signal's trend is fundamental. Without significant a priori information, however, recognizing the set of correct components in a signal of interest is practically impossible since the computational complexity renders the effort useless \cite{von2016adaptive}. In this sense, the hypothesis that the underlying trend has a sparse representation is a vital condition for practical implementations, which, in turn, leads to a similar problem since the counting problem associated to finding the most sparse representation of a signal also yields to computationally intractable problems. For that reason, the $\ell_1$ norm relaxation has become a universal proxy for the sparsity in recent years. The seminal work of Tibshirani on the \textit{lasso} \cite{tibshirani1996regression} has exposed the potential of a simple convex relaxation as the standard tool when dealing with high-dimensional data. Contemporaneously, Chen, Donoho and Saunders presented the \textit{basis pursuit} \cite{chen1994basis, chen2001atomic}, a very similar method in the context of sparse signal representation in over-complete dictionaries. These works, associated with some theoretical results \cite{donoho1989uncertainty, tropp2006just}, have led to the development of \textit{compressed sensing}\cite{donoho2006compressed, candes2006robust} and the improvement of existing fields such as machine learning, statistics and control.

Following the success of the lasso, a number of extensions with a variety of properties were proposed: the \textit{elastic net} \cite{zou2005regularization}, \textit{group lasso}\cite{yuan2006model} and the \textit{sparse-group lasso} \cite{simon2013sparse}, the \textit{graphical lasso} \cite{friedman2008sparse}, \textit{sparse PCA} \cite{zou2006sparse} and \textit{sparse clustering} \cite{elhamifar2013sparse}. Among this diversity of generalizations, we are particularly interested in the \textit{adaptive lasso} proposed by Zou \cite{zou2006adaptive}. The adaptive lasso uses appropriate weights for penalizing the coefficients based on a previous estimate which can be given by any consistent estimate, e.g., the least-squares solution. The main advantage of this methodology is the consistency in variable selection. In other words, as the number of observations increases, the underlying true model should be successfully identified.

Besides identifying the components of the underlying signal's trend, one is often interested in detecting abrupt level shifts which can impart useful information about the signal's characteristics. There are methods based on searching the signal for observations that can be considered a significant level shift, which are usually accompanied by some statistical test to validate the presence of such level-shift \cite{chow1960tests,quandt1960tests,brown1975techniques}. An efficient technique for searching potential structural breaks is via the Atheoretical Regression Tree (ART) \cite{rea2010identification}, in which the signal is sequentially split in the search for level-shifts. One of the most recent methods for identifying level shifts is through the minimization of the Potts functional \cite{storath2014jump}. 

The \textit{$\ell_1$ Trend Filter} \cite{kim2009ell_1} successfully tackles the problem of jointly estimating the trend of a time-series and identifying potential abrupt changes at unknown times. This is accomplished by assuming that the presence of level-shifts and outliers is sparse and that the signal can be approximated by piece-wise linear trends. In this document, we show that sparsity can be exploited in the $\ell_1$ Trend Filter by combining it with the adaptive lasso weighted penalty and enforcing multiple trend estimation in the form of a trend decomposition. We also propose a custom coordinate descent algorithm for the novel filter based on the \textit{covariance update method} \cite{friedman2007pathwise}.

\section{$\ell_1$ Adaptive Trend Filter}

In its traditional form, the $\ell_1$ Adaptive Trend Filter is written as a minimization problem comprised of a loss-function and a set of regularizers. In this context, the loss-function is expressed by the square of the Euclidean norm and the adaptive regularizers $r_i: \mathbb{R} \times \mathbb{R}  \to \mathbb{R}$ are based on the $\ell_1$-norm, which reads 
\begin{align*}
&\mu^{alt}_{(\lambda, \gamma)} = \text{arg}\min_{(\mu)} \tfrac{1}{2}||y - \mu||^2_2 + \lambda \sum_{i=1}^{n} r_i(\mu_i, \gamma).
\end{align*}
The main idea is to both select and estimate the trend's components that best fits the signal $y \in \mathbb{R}^n$ in a parsimonious manner. As in the adaptive lasso, the $\ell_1$ penalty is divided by an initial estimate for $\mu$. Even though any consistent estimator is suitable according to \cite{zou2006adaptive}, we focus on the entry-wise ordinary least squares (ols) estimator due to its simplicity and effectiveness. It has been shown in \cite{zou2006adaptive} that the use of an initial estimate as a weight for the regularization guarantees oracle properties for feature selection under mild conditions. In other words, the variable selection works asymptotically as if the correct set of true components were previously known.

We are going to decompose the trend as $\mu = x + w + u + s$, where $x$ corresponds to a piecewise linear trend, $w$ represents the level of the signal, $u$ is the outlier -- or spike -- component and $s$ is a seasonal component. Each of these components has a particular characterization, as well as an associated regularizer, which is described as follows:
\begin{itemize}
\item To induce the trend to have as fewer linear segments as possible the second difference of the piecewise linear trend is penalized as:
\begin{equation*}
r_x(x_i, \gamma) =
\begin{cases}
\frac{|x_{i-1} - 2 x_i + x_{i+1}|}{|x_{i-1}^{ols} - 2 x_i^{ols} +x_{i+1}^{ols}|^{\gamma}}, & \text{if $i \in [2, n-1]$}\\
0, & \text{otherwise.}
\end{cases}
\end{equation*}
\item In order to stimulate sparsity of level-shifts, the following adaptive regularizer is defined:
\begin{equation*}
r_w(w_i, \gamma) =
\begin{cases}
\frac{|w_i - w_{i-1}|}{|w_i^{ols} - w_{i-1}^{ols}|^{\gamma}}, & \text{if $i \in [2, n]$}\\
0, & \text{otherwise.}
\end{cases}
\end{equation*}
\item The subsequent regularizer is set to avoid the filter to overfill the trend with outliers:
\begin{align*}
&r_u(u_i, \gamma) = \frac{|u_i|}{|u_i^{ols}|^{\gamma}}.
\end{align*}
\item The filter is charged with the responsibility of selecting the best frequencies from an over-complete dictionary $\Omega$, where the frequency $\omega$ does not need to be known \textit{a priori}.
\begin{align*}
&s_i = \sum_{\omega \in \Omega}(a_{\omega} \sin{\omega i} + b_{\omega} \cos{\omega i}).
\end{align*}
To discourage a greedy use of sinusoidal components, it is enough to penalize their coefficients as: 
\begin{align*}
&r_s(s, \gamma) = \sum_{\omega \in \Omega}\frac{|a_{\omega}|}{|a_{\omega}^{ols}|^{\gamma}} + \frac{|b_{\omega}|}{|b_{\omega}^{ols}|^{\gamma}}.
\end{align*}
\end{itemize}

It is important to highlight that the estimate $\mu^{alt}_{(\lambda, \gamma)}$ is highly dependent on previously chosen penalizers $\lambda$ and $\gamma$ which work as control knobs for the degree of sparsity imposed by the filter. In practice, the estimate $\mu^{alt}_{(\lambda, \gamma)}$ is evaluated in a two-dimensional grid $(\lambda, \gamma) \in \Lambda \times \Gamma$ which is then selected according to an information criteria such as AIC \cite{akaike1987factor}, BIC \cite{bhat2010derivation}, Mallow's $C_p$ \cite{mallows1973some}, EBIC \cite{chen2008extended} or cross-validation \cite{kohavi1995study}. The choice of EBIC (Extended Bayesian Information Criteria) is more suitable for the application of this paper since it is designed to deal with high-dimensional complexity models \cite{chen2012extended}. Despite its popularity, cross-validation cannot be applied in a straightforward manner when samples are intrinsically ordered, as in the case of time-series.

One may also want to establish lower and upper bounds to a component, e.g., it can be useful to set all level-shifts to be non-negative. This kind of constraint can be directly embedded to the aforementioned filter in the form of a linear constraint without adding further computational complexity.

\section{Fast Covariance Updating Coordinate Descent Algorithm}

For the sake or clarity in the algorithm, we redefine the $\ell_1$ Adaptive Trend Filter in its \textit{matrix} form:
\begin{align*}
&\theta^{alt}_{(\lambda, \gamma)} = \text{arg}\min_{(\theta)} \tfrac{1}{2}||y - A \theta||^2_2 + \lambda \sum^{p}_{i=1}\frac{|\theta_i|}{|\theta^{ols}_i|^{\gamma}},
\end{align*}
where the \textit{traditional} and \textit{matrix} forms are linked by $\mu^{alt}_{(\lambda, \gamma)} = A\theta^{alt}_{(\lambda, \gamma)}$. In this framework, the matrix $A \in \mathbb{R}^{n \times p}$ works as an over-complete dictionary ($p>n$) of potential components to be chosen by the filter. The vector $\theta^{alt}_{(\lambda, \gamma)} \in \mathbb{R}^p$, expected to be sparse due to the $\ell_1$ adaptive regularization, defines the linear combination of the set of components that best fits the signal. The dictionary matrix can be seen as a concatenation of different matrices, one for each of the previously defined components of $\mu$, such that $A = [A^x \hspace{0.1cm} A^w \hspace{0.1cm} A^u \hspace{0.1cm} A^s \hspace{0.1cm} A^c]$, with:
\begin{align*}
&A^x\!=\!\begin{bmatrix*}
	0\!&\!0\!&\!\dots\!&\!0\\
	1\!&\!0\!&\!\dots\!&\!0\\
	\vdots\!&\!\vdots\!&\!\ddots\!&\!\vdots\\
	n\!-\!2\!&\!n\!-\!3\!&\!\dots\!&\!0\\
	n\!-\!1\!&\!n\!-\!2\!&\!\dots\!&\!1
\end{bmatrix*};\\
&A^w\!=\!\begin{bmatrix*}
	0\!&\!0\!&\!\dots\!&\!0\\
	1\!&\!0\!&\!\dots\!&\!0\\
	\vdots\!&\!\vdots\!&\!\ddots\!&\!\vdots\\
	1\!&\!1\!&\!\dots\!&\!0\\
	1\!&\!1\!&\!\dots\!&\!1
\end{bmatrix*};\hspace{0.5cm}
A^u\!=\!\begin{bmatrix*}
	1\!&\!0\!&\!\dots\!&\!0\\
	0\!&\!1\!&\!\dots\!&\!0\\
	\vdots\!&\!\vdots\!&\!\ddots\!&\!\vdots\\
	0\!&\!0\!&\!\dots\!&\!1
\end{bmatrix*};\\
&A^s\!=\!\begin{bmatrix*}
	|\!&\!|\!&\!\dots\!&\!|\\
	\text{\rotatebox{90}{$\sin\pars{\omega_1 t}$}}\!&\!\text{\rotatebox{90}{$\sin\pars{\omega_2 t}$}}& &\!\text{\rotatebox{90}{$\sin\pars{\omega_k t}$}}\\
	|\!&\!|\!&\!\dots\!&|
\end{bmatrix*};\hspace{0.08cm}
A^c\!=\!\begin{bmatrix*}
	|\!&\!|\!&\!\dots\!&\!|\\
	\text{\rotatebox{90}{$\cos\pars{\omega_1 t}$}}\!&\!\text{\rotatebox{90}{$\cos\pars{\omega_2 t}$}}& &\!\text{\rotatebox{90}{$\cos\pars{\omega_k t}$}}\\
	|\!&\!|\!&\!\dots\!&|
\end{bmatrix*}.
\end{align*}
It is worth emphasizing that our algorithm is a matrix-free method, i.e., it does not require the storage of the coefficients of $A$ since we are only interested in the inner products between the columns of $A$ (Refer to Alg. \ref{Algorithm}). Also, note that the number of columns on the matrices $A^s$ and $A^c$, as opposed to the other matrices, is related to the size of the over-complete set $\Omega$ and do not depend on the length of $y$.

The $\ell_1$ Adaptive Trend Filtering is comprised by two main parts: the convex and differentiable loss-function, represented by the Euclidean norm squared, and the convex and separable adaptive regularizer. As it has been explored in \cite{friedman2010regularization}, the coordinate descent method is particularly efficient to the lasso problem. The main idea behind the coordinate descent is very simple: the algorithm minimizes the multivariate objective function along one coordinate at a time by cyclically iterating through the coordinates until convergence. 

At any given iteration of the algorithm, the minimization is intended to explain the partial residual $r = y - \sum_{j \neq i}A_{j}\hat{\theta}_j$ -- the portion of the signal that was not explained by previous coordinates -- by solving a univariate optimization problem.  In the case of the lasso problem, the solution to such univariate optimization problem is analytical and is given by the \textit{soft-thresholding operator} \cite{donoho1995noising} of the inner product between the signal of interest and the column of $A$ associated with the active coordinate. The optimal solution is, thus, given by
\begin{align*}
&\hat{\theta}_{(\lambda, \gamma), i}^{alt} = \bigg(\tfrac{1}{\sigma_i^2}\bigg) \text{sign}\bigg(\langle\ r, A_i\rangle\bigg) \bigg(\frac{|\langle\ r, A_i\rangle|}{n}-\frac{\lambda}{|\theta^{ols}_i|^{\gamma}}\bigg),
\end{align*}
where $\langle\ r, A_i\rangle = \langle\ y, A_i\rangle - \sum_{k \neq i} \langle\ A_k, A_i\rangle \hat{\theta}_k^{alt}$.

The active set $\mathcal{A}$ is responsible for tracking the coefficients that are currently nonzero. For each tuple $(\lambda, \gamma)$, convergence is reached if the active-set does not change after an entire run through the $p$ coordinates. The grid $\Lambda \times \Gamma$ is evaluated in a decreasing order of $\lambda$ for any given $\gamma$ so the estimate from the previous cycle can be used as a \textit{warm-start}. We shall write the coordinate descent algorithm in its pseudo-code form as proposed by \cite{friedman2007pathwise}
\begin{algorithm}[H]
	\caption{Coordinate Descent}
	\label{Algorithm} 
	\begin{algorithmic}
		\STATE{Compute $\langle A_j , y \rangle \hspace{0.1cm} \forall \hspace{0.1cm} j=1, \dots, p$ }
		\FORALL{$(\lambda, \gamma) \in \Lambda \times \Gamma$}
		\STATE{$\mathcal{A} = \varnothing$}
		\WHILE{$\mathcal{A}$ did not converge}
		\FORALL{$i \in \{1,.., p\}$}
		\STATE{$\langle\ r, A_i\rangle = \langle A_i , y \rangle - \displaystyle\sum_{j \in \mathcal{A}: j \neq i}\langle A_i , A_j \rangle \hspace{0.05cm} \hat{\theta}_{(\lambda, \gamma), j}^{alt}$}
		\STATE{$\hat{\theta}_{(\lambda, \gamma), i}^{alt} = \tfrac{\text{sign}(\langle\ r, A_i\rangle)}{\sigma_i^2} \bigg(\frac{|\langle\ r, A_i\rangle|}{n}-\frac{\lambda}{|\theta^{ols}_i|^{\gamma}}\bigg)_{+}$}
		\IF{$|\hat{\theta}_{(\lambda, \gamma), i}^{alt}| > 0$ \textbf{and} $j \notin \mathcal{A}$}
		\STATE{$\mathcal{A} = \mathcal{A} \cup \{ j\}$}
		\ENDIF
		\ENDFOR
		\ENDWHILE
		\ENDFOR
	\end{algorithmic}
\end{algorithm}
The complexity of the algorithm resides on the number of runs through the $p$ coordinates and on the inner products between the columns of $A$, which have an associated cost of $\mathcal{O}(n^2)$. However, instead of directly computing the inner products, one can benefit from the particular design of matrix $A$ and establish analytic formulas for the inner products between pairs of columns to achieve a computational cost of $\mathcal{O}(1)$. This modification reduces the most expensive step of the algorithm, which will result in a faster implementation of the $\ell_1$ Adaptive Trend Filter. Such formulas are available at \cite{GitJuliaImplementation}.

Supposing the algorithm is at a particular iteration where $k$ features are present in $\mathcal{A}$, every cycle costs $\mathcal{O}(pk)$ operations despite any possible updates in the active-set, as opposed to the traditional coordinate descent method presented in \cite{friedman2007pathwise},in which the entry of each new feature into the active set requires an additional of $\mathcal{O}(nk)$ computations. This property is specially advantageous for smaller values of $(\lambda, \gamma)$ when more features tend to be included in $\mathcal{A}$.

\section{Applications}

\subsection{Optical Fiber Fault Detection}

Despite the great efficiency of fiber optics in transmitting data, they are made of very fragile material in which small torsions and compressions can result in mechanical ruptures. As a consequence, data transmission might be compromised or even interrupted. Since optical fibers links are typically longer than two kilometers, finding the location of the rupture with a maximal accuracy is operationally useful.

Optical Time Domain Reflectometry (OTDR) offers a method for detecting and evaluating fiber faults without the need of end-to-end measurements thus providing a valuable tool to support the operation of Fiber-Optic links \cite{BarnoskiAO1977}. A raw OTDR fiber signature is usually presented as the optical power at the detector in logarithmic scale versus the distance, for in that scale it is represented as a descending line with slope equal to the fiber's attenuation coefficient \cite{AmaralJLT2015}. In this context, the linear trend reflects the power attenuation along the fiber, the step trend indicates the presence of optical losses such as splitters, connectors or fiber defects, the spike trend indicates the presence of reflective events and the noise, which should be filtered out from the original series, may represent a number of sources of imprecision during data acquisition \cite{AmaralJLT2015}\cite{von2016adaptive}.

\begin{figure}[ht]
\centering
\includegraphics[width=0.95\linewidth]{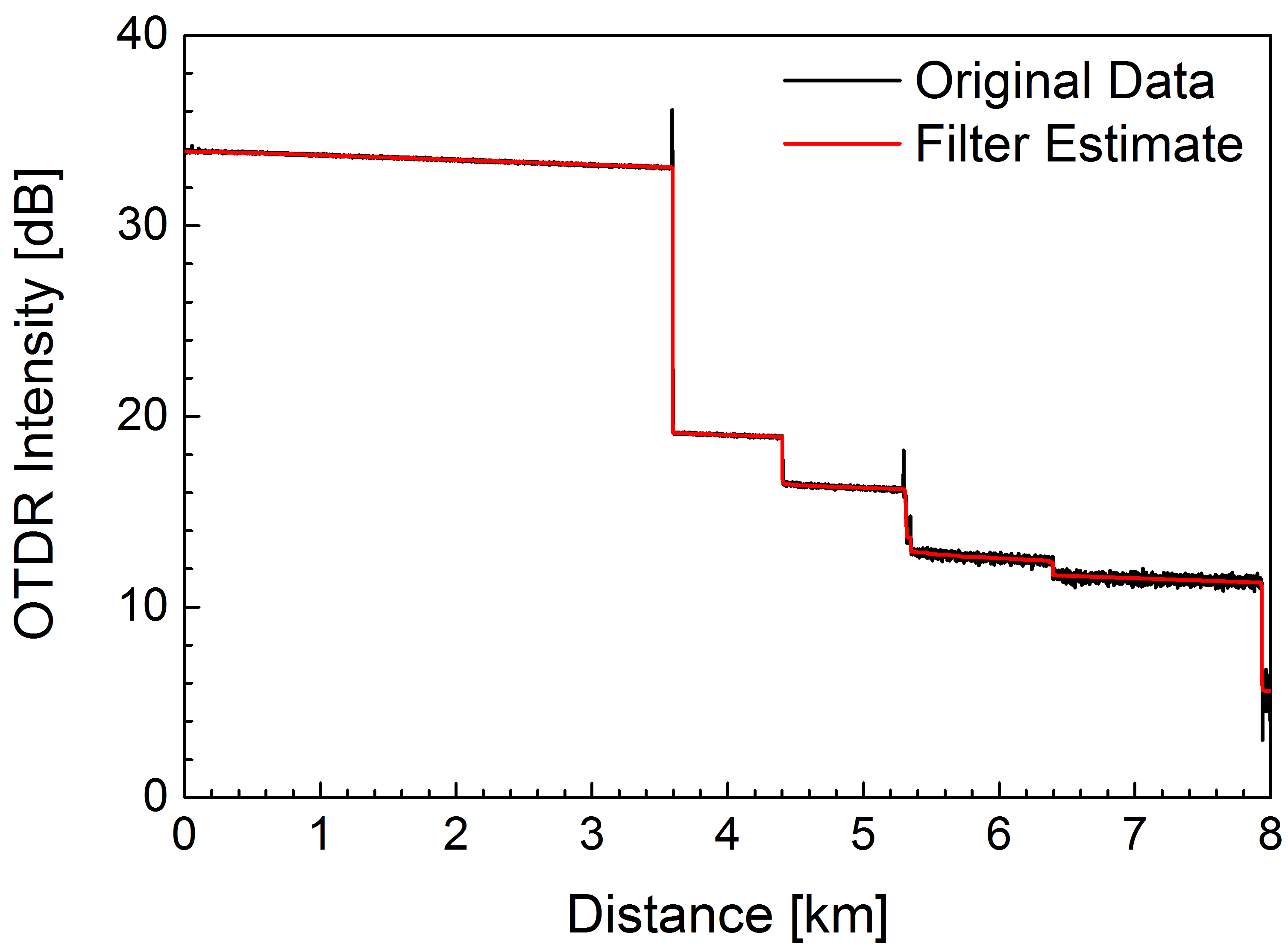}
\caption{$\ell 1$ Adaptive Filter result for an OTDR fiber profile. The filter is able to identify the components that constitute the data series, which, in this specific application, is equivalent to automatically identifying the fiber fault locations in the test-bench optical fiber link. Note that the sinusoidal contribution to the overall result was negligible, attesting that the filter is highly discriminatory in terms of the contributions to its estimate.}
\label{fig:PON_case_study}
\end{figure}

In this case study, we consider an $8$-kilometers point-to-multipoint optical fiber communication link in which a $\sim3$-kilometers feeder fiber is directed to a $1\times32$ passive optical splitter representing the remote distribution node. The goal is to be able to identify any kind of losses -- both small ($\leq1.0$ dB) and high ($\geq1.0$ dB) -- after the high-splitting ratio device \cite{YnoquioOFC2016}. In Fig.\ref{fig:PON_case_study}, we present the original and filtered data series when the $\ell 1$ Adaptive Filter is employed.

The elimination of the noise component and the accurate description of the signal in terms of steps and spikes is clear, i.e., the $\ell_1$ adaptive filter selected the best components to fit the observed signal.

\subsection{Wind Farm Power Generation}

The main challenge of integrating wind power into existing systems is the volatile nature of wind power generation. In this sense, one needs a forecasting model that takes into account the seasonal and intermittent behaviour of wind farms. This task is complicated by the presence of the two following reasons. Firstly, unscheduled periods of maintenance and repair often produce long sequences of zero power generation in the data. Simple repairs, like electrical equipment failure, are usually resolved in the range of one to three days but there might be more complicated issues where the maintenance time can be extended for weeks. Secondly, wind turbines are designed with \textit{cut-in speed}, i.e. the minimum wind speed at which the turbine will generate power, and \textit{cut-out speed}, i.e. the wind speed at which the turbine is shut down for safety reasons.

\begin{figure}[ht]
\centering
\includegraphics[width=0.95\linewidth]{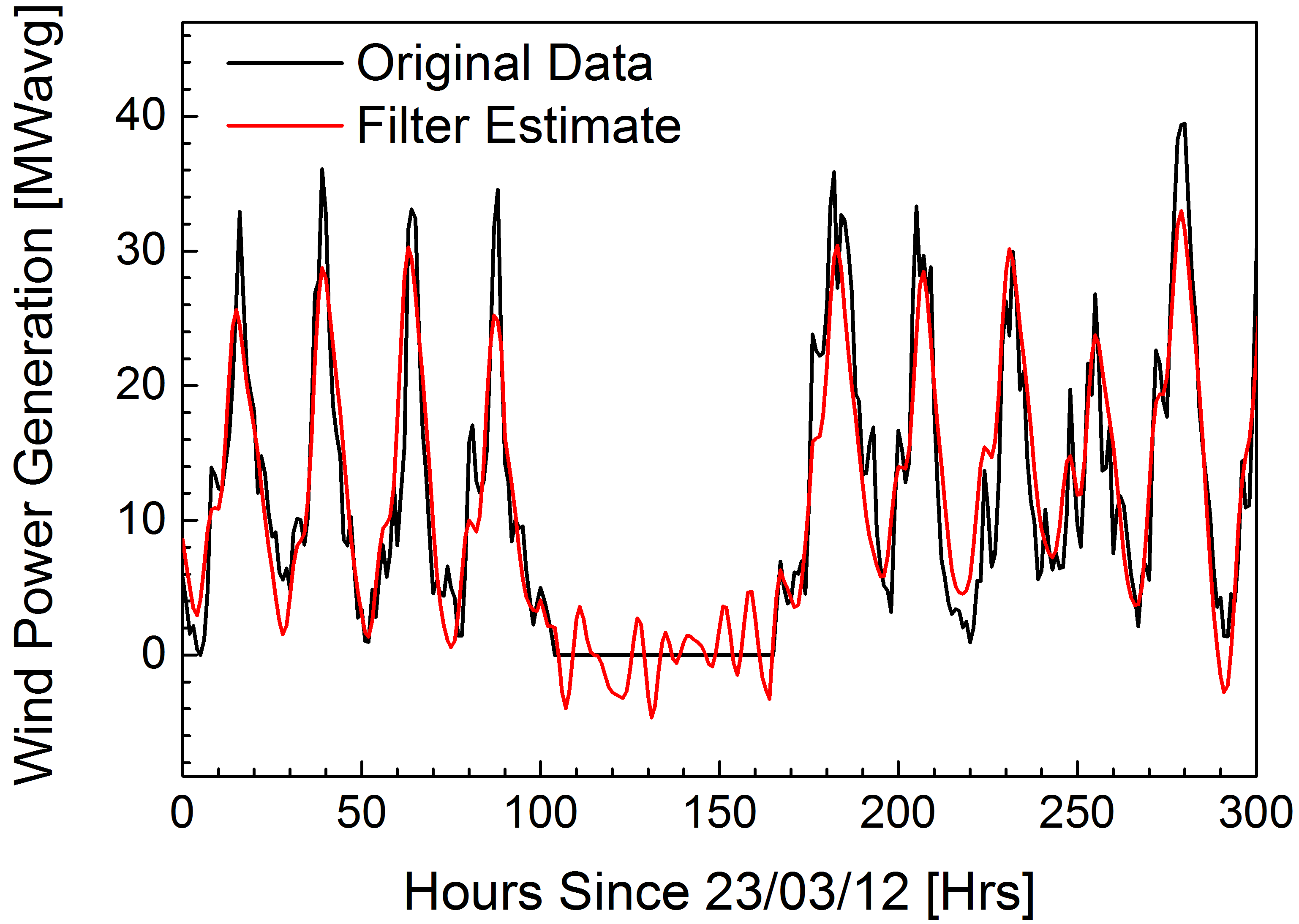}
\caption{$\ell 1$ Adaptive Filter result for a set of fourteen-day hourly measurements of power generation from a Brazilian wind farm. The filter is able to identify the components that constitute the data series. Note that a level-shift contribution has been added to account for the wind farm shutdown period.}
\label{fig:wind_case_study}
\end{figure}

In this case study, we consider a set of fourteen-day hourly measurements of power generation from a Brazilian wind farm. Each measurement is the mean of power generated in the respective hour in megawatts. It is previously known that maintenance takes place between the 27th and the 29th of March. Nevertheless, this information is not given to the filter. As can be seen in Fig.\ref{fig:wind_case_study}, the hourly power generation is definitely periodic: energy production is very low by the earliest hours of the day; and the generation peak is achieved in the evening. 

From the filter's estimate result, we see that no slope or spike components were selected. The estimated trend is comprised mainly by trigonometric features, as it was expected due the presence of strong periodicity. An over-complete set $\Omega$ of forty two potential frequencies were given to the filter, ranging from a periodicity of six to forty eight hours. The $\ell_1$ adaptive filter selected an amount of three cosines components and seven sines. The most important frequency, i.e., the one associated with the largest absolute value amplitude, corresponds to the twenty four hours periodicity as it would be expected from a daily-periodic signal.

Through the step component, the filter was able to accurately identifying the stage of maintenance. With a difference of one hour at the beginning and two hours at the end, a cluster of three consecutive negative steps followed by a cluster of positive steps represents the abrupt change in the wind power production. This kind of analysis has two main advantages: firstly, one can automatize the identification of potential failures and maintenance track record; secondly, the presence of step and spike components avoid anomalous observations to be erroneously embedded in the trigonometric features.

In this brief study, it is shown that the $\ell_1$ adaptive filter can correctly identify and fit in-sample wind power data without the need of any data pre-processing techniques. In future work, it is important to investigate the potential of the filter as a forecasting tool in comparison to existing technologies \cite{huang1995use}\cite{souto2014high}\cite{larsen2016joint}. This feature has the potential for predicting renewable energy loads in electrical power grids. Considering the seasonal behaviour of other renewable resources. The authors believe that the same method can be applied to solar and hydro-based power generation.

\section{Conclusion}
In this paper, we proposed the $\ell_1$ adaptive filter as signal processing tool able to consistently identify a sparse set of components comprising the trend of the signal of interest. Some advantages of this methodology are: robustness to outliers, lack of tuning parameters or data pre-processing. 

An extensive literature in enhancing the coordinate descent for the $\ell_1$-regularized loss minimization is available. In this sense, potential improvements in the Julia implementation needs to be investigated. In special, the use of strong rules \cite{tibshirani2012strong}, randomization \cite{fercoq2015accelerated}\cite{richtarik2014iteration} and parallelization \cite{bradley2011parallel}\cite{richtarik2015parallel}.

\section*{Acknowledgment}

The authors would like to thank Brazilian Agencies CAPES, CNPq and FAPERJ for financial support.

\section{Supplemental Material}

The authors provide the complete Julia implementation of the herewith presented $\ell_1$ Adaptive Filter as a digital supplemental material in the "L1AdaptiveTrendFilter.jl" package \cite{GitJuliaImplementation}. The code is open source and available through the MIT license. We hope this can foster discussions on how to improve the implementation in order to make it more useful for the signal processing community.

\bibliographystyle{IEEEtran}
\bibliography{Ref}
\vfill

\end{document}